\documentstyle[12pt,epsf]{article}
\renewcommand{\bar}[1]{\overline{#1}}

\begin{document}

\begin{flushright}
USM-TH-81
//hep-ph/9907556
\end{flushright}

\bigskip\bigskip
\centerline{\large \bf Flavor and Spin Structure of Octet
Baryons at Large x}

\vspace{22pt}

\centerline{\bf Bo-Qiang Ma$^{a}$, Ivan Schmidt$^{b}$, and
Jian-Jun Yang$^{b,c}$}

\vspace{8pt} 
{\centerline {$^{a}$Department of Physics, Peking University,
Beijing 100871, China,\footnote{Mailing address}}

{\centerline {CCAST (World Laboratory),
P.O.~Box 8730, Beijing 100080, China,}}

{\centerline {and Institute of High Energy Physics, Academia
Sinica, P.~O.~Box 918(4),}}

{\centerline {Beijing 100039, China}}

{\centerline{e-mail: mabq@phy.pku.edu.cn}}

{\centerline {$^{b}$Departamento de F\'\i sica, Universidad
T\'ecnica Federico Santa Mar\'\i a,}}

{\centerline {Casilla 110-V, 
Valpara\'\i so, Chile}}

{\centerline{Email: ischmidt@fis.utfsm.cl} }

{\centerline {$^{c}$Department of Physics, Nanjing Normal
University,}}

{\centerline {Nanjing 210097, China}}

{\centerline{Email: jjyang@fis.utfsm.cl} }


\vspace{10pt}
\begin{center} {\large \bf Abstract}

\end{center}

The quark flavor and spin distributions in octet baryons are
calculated both in the SU(6) quark spectator diquark model and in
a perturbative QCD (pQCD) based model. It is shown that the
$\Lambda$ has the most significant difference in flavor structure
at large $x$  between the two models, though the flavor and spin
structure of other baryons can also provide tests of different
models. The Drell-Yan process for $\Sigma^{\pm}$ beams on
isoscalar targets can be used to test different predictions
concerning the valence quark flavor structure of the
$\Sigma^{\pm}$.

\vfill

\centerline{PACS numbers: 14.20.-c, 12.38.Bx, 12.39.Ki, 13.85.-t}

\vfill
\centerline{Published in Nucl.~Phys.~B 574 (2000) 331 }
\vfill
\newpage

\section{Introduction}

Parton distributions of hadrons and the formation of hadrons from
fragmentation of partons are of considerable current interest in
the community of particle and nuclear physics. There have been
remarkable achievements in our knowledge of the quark-gluon
structure of the nucleon from three decades of experimental and
theoretical investigations in various deep inelastic scattering
(DIS) processes. However, there are still a number of unknowns
concerning the detailed flavor and spin structure of the nucleon,
such as the detailed origin of the proton spin \cite{SpinR}, the
strange content of the nucleon \cite{Bro88,Bro96}, the flavor
asymmetry of the sea \cite{Kum97}, the isospin symmetry breaking
at small $x$ \cite{Isospin}, and the flavor and spin structure of
the valence quarks at large $x$
\cite{Far75,DQM,Bro95,Ma96,Mel96,Yang99}. It is important to
perform high precision measurements of available physical
quantities and/or to measure new quantities related to the flavor
and spin structure of the nucleons, in order to have a better
understanding of the quark-gluon structure of the nucleon.
Nevertheless,
it has recently been  found \cite{MSY2} that the flavor
and spin structure of the $\Lambda$-hyperon may serve as a new
domain where the same physics that governs the structure of the
nucleon can manifest itself. It was  found that the flavor and
spin structure of the quark distributions of the $\Lambda$ differ
significantly at large $x$ from a perturbative QCD (pQCD) based
analysis and a SU(6) quark-diquark model. A detailed analysis
\cite{MSY3} of the available $\Lambda$-polarization data in
$e^+e^-$ annihilation at the $Z$-pole supports the prediction that
the $u$ and $d$ quarks inside the $\Lambda$ should be positively
polarized at large $x$, though their net helicities might be zero
or negative.  The most recent HERMES result \cite{HERMES99b}
of spin transfer
to the $\Lambda$ in deep elastic scattering of polarized
lepton on the nucleon target also support the predictions
of the SU(6) quark-diquark model and the pQCD based model
\cite{MSY2}.

The direct measurements of the $\Lambda$ quark structure are not
easy since it is a charged neutral particle which cannot be
accelerated as incident beam and its short life time makes it also
difficult to be used as target. The simple reciprocity relation
\cite{GLR}
\begin{equation}
q_{h}(x) \propto D_q^h(z),
\end{equation}
where $z=2 p \cdot q/Q^2$ is the momentum fraction of the produced
hadron from the quark jet in the fragmentation process and
$x=Q^2/2 p \cdot q$ is the Bjorken scaling variable corresponding
to the momentum fraction of the quark from the hadron in the DIS
process, can provide a reasonable connection between different
physical quantities and lead to understandings of the $\Lambda$
quark structure from the various quark to $\Lambda$
fragmentations. However, such a relation is still not completely
free from theoretical and experimentally uncertainties, though it
may  serve for our purpose as an approximate qualitative
connection at a specific scale $Q^2$, near $x \to 1$ and $z \to 1$
\cite{Bro97,Ma99}. Thus the direct measurement of the quark
structure of other octet baryons other than the nucleon and
$\Lambda$ has a strong physical significance, and can provide a
new direction to test different theories concerning the nucleon
structure.

The purpose of this paper is to extend the analysis of the quark
structure from the nucleon and $\Lambda$ cases to the other
members of octet baryons. We will calculate the quark
distributions for all of the octet baryons in the SU(6) quark
spectator diquark model and in a perturbative QCD based model.
There are two motivations for such a study: (1) to check the
difference in flavor and spin structure of all octet baryons
between the two models, and see which baryon has the most
significant difference; (2) some  charged baryons other than
nucleons, such as $\Sigma^{\pm}$, may be used as beam to directly
measure their own quark structure in case the structure of the
target is comparatively well known. We will find that the
$\Lambda$ has the most significant difference at large $x$ in the
flavor and spin structure for a clean test of different
predictions. We will also show that the $\Sigma$'s have the most
significant difference in the flavor and spin structure between
the two models at medium to large $x$, and the measurement of Drell-Yan
process for $\Sigma^{\pm}$ beams on the isoscalar targets can test
different predictions of the quark structure of the $\Sigma^{\pm}$
baryons. It is more appealing that
the $\Lambda$ and $\Sigma^0$ have complete different flavor and spin
structure though they are composed of same flavor quarks.

We shall start Sec.~II with the presentation of quark
distributions of octet baryons in  the SU(6) quark-diquark
spectator model and in a perturbative QCD based model. We then
compare the flavor and spin structure of all the octet baryons
between the two models at large $x$. Then in Sec.~III, we present
the formulas for the Drell-Yan process using $\Sigma^{\pm}$ beams
on isoscalar targets and show different predictions of cross
section ratios in the pQCD based model and in the quark spectator
diquark model. Finally, we present conclusions in Section IV.

\section{The Quark Spin and Flavor Structure of  Octet Baryons}

In this section, we extend the analysis of the quark structure of
nucleons to all members of octet baryons using (1) the SU(6)
quark spectator diquark model and, (2) a perturbative QCD (pQCD)
based model.

\subsection{SU(6) Quark Spectator Diquark Model}

Before we look into the details of the  spin and the flavor
structure for the valence quarks of the octet baryons, we briefly
review the analysis of the unpolarized and polarized quark
distributions of nucleons in the light-cone SU(6)
quark-spectator-diquark model \cite{Ma96}.
As we know, it is proper to
describe deep inelastic scattering as the sum of incoherent
scatterings of the incident lepton on the partons in the infinite
momentum frame or in the light-cone formalism. The unpolarized
valence quark distributions $u_v(x)$ and $d_v(x)$ of the proton
are given in this model by
\begin{eqnarray}
&&u_{v}(x)=\frac{1}{2}a_S(x)+\frac{1}{6}a_V(x);\nonumber\\
&&d_{v}(x)=\frac{1}{3}a_V(x), \label{eq:ud}
\end{eqnarray}
where $a_D(x)$ ($D=S$ for scalar spectator or $V$ for axial vector
spectator) can be expressed  in terms of the light-cone momentum
space wave function $\varphi (x, \vec{k}_\perp)$ as
\begin{equation}
a_{D}(x) \propto  \int [\rm{d}^2 \vec{k}_\perp] |\varphi (x,
\vec{k}_\perp)|^2, \hspace{1cm} (D=S \hspace{0.2cm} or
\hspace{0.2cm} V)
\end{equation}
which is normalized such that $\int_0^1 {\mathrm d} x a_D(x)=3$
and denotes the amplitude for quark $q$ to be scattered while the
spectator is in the diquark state $D$.

The quark helicity distributions for the $u$ and $d$ quarks in the
proton  can be written as \cite{Ma96}
\begin{equation}
\begin{array}{llcr}
\Delta u_{v}(x)=u_{v}^{\uparrow}(x)-u_{v}^{\downarrow}(x)
=-\frac{1}{18}a_V(x)W_V(x) +\frac{1}{2}a_S(x)W_S(x);
\\
\Delta d_{v}(x)=d_{v}^{\uparrow}(x)-d_{v}^{\downarrow}(x)
=-\frac{1}{9}a_V(x)W_V(x),
\end{array}
\label{eq:sfdud}
\end{equation}
in which $W_V(x)$ and $W_S(x)$ are the Melosh-Wigner correction
factors \cite{Ma91b} for the  axial vector  and scalar
spectator-diquark cases. They are obtained by averaging
\begin{equation}
W_D(x,{\mathbf k}_{\perp}) =\frac{(k^+ +m_q)^2-{\mathbf
k}^2_{\perp}} {(k^+ +m_q)^2+{\mathbf k}^2_{\perp}} \label{eqM1},
\end{equation}
over ${\mathbf k}_{\perp}$ with $k^+=x {\cal M}$ and ${\cal
M}^2=\frac{m^2_q+{\mathbf k}^2_{\perp}}{x}+\frac{m^2_D+{\mathbf
k}^2_{\perp}}{1-x}$, where $m_D$ is the mass of the diquark
spectator, and are unequal due to unequal spectator masses, which
leads to unequal ${\mathbf k}_{\perp}$ distributions.

Now, we extend the above analysis of the valence quark
distributions of the proton to that of the octet baryons. The
valence quark distributions of the $\Lambda$ in the SU(6) quark
spectator diquark model have been analyzed in detail in
Ref.\cite{MSY3}. The $\Lambda$ wave function in the conventional
SU(6) quark model is written as
\begin{equation}
|\Lambda^{\uparrow} \rangle =\frac{1}{2\sqrt{3}} [(u^{\uparrow} d
^{\downarrow} + d^{\downarrow} u ^{\uparrow}) -(u^{\downarrow} d
^{\uparrow} + d ^{\uparrow} u ^{\downarrow} )] s^{\uparrow} +
(\mathrm{cyclic ~~permutation}). \label{SU6}
\end{equation}
The SU(6) quark-diquark model wave function for the $\Lambda$ is
written as
\begin{equation}
\Psi^{\uparrow,\downarrow}_{\Lambda} =\sin \theta ~ \varphi_V |q V
\rangle_{\Lambda} ^{\uparrow,\downarrow} + \cos \theta ~ \varphi_S
|q S \rangle_{\Lambda}^{\uparrow,\downarrow}, \label{SU6L}
\end{equation}
with
\begin{equation}
\begin{array}{cllr}
|q V \rangle_{\Lambda} ^{\uparrow,\downarrow}&=&\pm
 \frac{1}{\sqrt{6}} [V_0(ds) u
^{\uparrow,\downarrow} - V_0(us) d ^{\uparrow,\downarrow} -
\sqrt{2} V_{\pm}(ds) u ^{\downarrow,\uparrow} + \sqrt{2}
V_{\pm}(us) d ^{\downarrow,\uparrow}];
\\
|q S \rangle_{\Lambda} ^{\uparrow,\downarrow}&=&\frac{1}{\sqrt{6}}
[S(ds) u ^{\uparrow,\downarrow} + S(us) d ^{\uparrow,\downarrow}
-2  S(ud) s ^{\uparrow,\downarrow}],
\end{array}
\label{SU6LD}
\end{equation}
where $V_{s_z}(q_1 q_2)$ stands for a $q_1 q_2$ vector diquark
Fock state with third spin component $s_z$, $S(q_1 q_2)$ stands
for a $q_1q_2$ scalar diquark Fock state, and $\varphi_D$ stands
for the momentum space wave function of the quark-diquark with $D$
representing the vector (V) or scalar (S) diquarks. The angle
$\theta$ is a mixing angle that breaks the SU(6) symmetry at
$\theta \neq \pi/4$ and in this paper we choose the bulk SU(6)
symmetry case $\theta =\pi/4$.

We analyze the valence quark distributions of the $\Sigma^0$ by
extending the SU(6) quark-spectator-diquark model from the nucleon
\cite{Ma96} and $\Lambda$ \cite{MSY3} cases to the $\Sigma^0$.
Similarly, the $\Sigma^0$ wave function in the conventional SU(6)
quark model is

\begin{eqnarray}
|\Sigma^{0\uparrow} \rangle &=& \frac{1}{3} (u^{\uparrow} d
^{\uparrow} + d^{\uparrow} u ^{\uparrow})s^{\downarrow}\nonumber\\
 &-&\frac{1}{6}(u^{\uparrow}d ^{\downarrow}+ u ^{\downarrow}d^{\uparrow}
 +d^{\uparrow}u^{\downarrow}+d^{\downarrow}u^{\uparrow} ) s^{\uparrow} +
(\mathrm{cyclic ~~permutation}). \label{SU6s}
\end{eqnarray}

The SU(6) quark-diquark model wave function for the $\Sigma^0$ is
written as
\begin{equation}
\Psi^{\uparrow,\downarrow}_{\Sigma^0} =\sin \theta ~ \varphi_V |q
V \rangle_{\Sigma^0}^{\uparrow,\downarrow} + \cos \theta ~
\varphi_S |q S \rangle_{\Sigma^0}^{\uparrow,\downarrow},
\label{SU6Ls}
\end{equation}
with
\begin{eqnarray}
|q V \rangle_{\Sigma^0} ^{\uparrow,\downarrow}&=&\pm
 \frac{1}{3} [\sqrt{2} V_0(ud) s
^{\uparrow,\downarrow} - \frac{1}{\sqrt{2}} V_0(us) d
^{\uparrow,\downarrow} - \frac{1}{\sqrt{2}} V_0(ds) u
^{\uparrow,\downarrow} \\\nonumber &-&2 V_{\pm}(ud) s
^{\downarrow,\uparrow} + V_{\pm}(us) d ^{\downarrow,\uparrow}+
V_{\pm}(ds) u ^{\downarrow,\uparrow}];
\end{eqnarray}

\begin{equation}
|q S \rangle_{\Sigma^0} ^{\uparrow,\downarrow} =
-\frac{1}{\sqrt{2}} [S(us) d ^{\uparrow,\downarrow} + S(ds) u
^{\uparrow,\downarrow}].
\end{equation}

Instead of writing the wave functions for other octet baryons as
above, in Tab.~1 we present  all quark distributions of octet
baryons in SU(6) quark spectator diquark model with the
quark-diquark amplitude $a_{D}$ with $D=V$, $S$.
 The Melosh-Wigner rotation
effect in the quark spin distributions is denoted by the amplitude
\begin{equation}
\tilde{a}_{D}=a_{D}(x)W_{D}(x),
\end{equation}

 In  the calculation,
we employ the Brodsky-Huang-Lepage (BHL) prescription \cite{BHL}
of the light-cone momentum space wave function for the
quark-spectator
\begin{equation}
\varphi (x, \vec{k}_\perp) = A_D \exp \{-\frac{1}{8\alpha_D^2}
[\frac{m_q^2+\vec{k}_\perp ^2}{x} +
\frac{m_D^2+\vec{k}_\perp^2}{1-x}]\},
\end{equation}
with the parameter $\alpha_D=330$. Other parameters such as the
quark mass $m_q$, vector(scalar) diquark mass $m_{V(S)}$ for
baryons of the octet are listed in the table.

The quark structure of the other members of the octet baryons are
connected to the proton by SU(3) symmetry with the valence quarks
$q_1=u$ and $q_2=d$ for proton replaced by $q_1$ and $q_2$ in
Tab.~2 for the corresponding baryon. In the quark spectator
diquark model the exact SU(3) symmetry is broken due to the mass
difference for different quarks and diquarks, as shown in Tab.1.

\vspace{0.5cm}
\newpage

\centerline{Table~1~~ The quark distribution functions of octet
baryons in SU(6) quark-diquark model}

\vspace{0.3cm}

\begin{footnotesize}
\begin{center}
\begin{tabular}{|c||c|c||c|c||c|c|c|}\hline
 ~~~~~~~~~~~~~~~& $~~~~~~~$ & ~~~~~~~~~~~~~~~& $~~~~~~$ &~~~~~~~~~~~~~~~&
$m_q$ & $m_V$ & $m_S$
\\
Baryon& $~~q ~~$ & ~~~~~~~~~~~~~~~& $~~\Delta q ~~$
&~~~~~~~~~~~~~~~& (MeV) & (MeV) & (MeV)
\\ \hline
 ~~~~p~~~~ & $~~u~~$ &$\frac{1}{6}a_V+\frac{1}{2}a_S $ &
 $\Delta u$ & -$\frac{1}{18}\tilde{a}_V+\frac{1}{2}\tilde{a}_S $ &
330 & 800 & 600 \\ \cline{2-8}
 (uud)& $~~d~~$ &$\frac{1}{3}a_V$ &
 $\Delta d$ & -$\frac{1}{9}\tilde{a}_V$ &
330 & 800 & 600 \\ \cline{1-8}
 ~~~~n~~~~ & $~~u~~$ &$\frac{1}{3}a_V $ &
 $\Delta u$ & -$\frac{1}{9}\tilde{a}_V $ &
330 & 800 & 600 \\ \cline{2-8}
 (udd)& $~~d~~$ &$\frac{1}{6}a_V+\frac{1}{2} a_S$ &
 $\Delta d$ & -$\frac{1}{18}\tilde{a}_V+\frac{1}{2}\tilde{a}_S$ &
330 & 800 & 600 \\ \cline{1-8}
 $~~~~\Sigma^{+}~~~~$ & $~~u~~$ &$\frac{1}{6}a_V+\frac{1}{2}a_S $ &
 $\Delta u$ & -$\frac{1}{18}\tilde{a}_V+\frac{1}{2}\tilde{a}_S $ &
330 & 950 & 750 \\ \cline{2-8}
 (uus)& $~~s~~$ &$\frac{1}{3}a_V$ &
 $\Delta s$ & -$\frac{1}{9}\tilde{a}_V$ &
480 & 800 & 600 \\ \cline{1-8}
 $~~~~\Sigma^{0}~~~~$ & $~~u~~$ &$\frac{1}{12}a_V+\frac{1}{4}a_S $ &
 $\Delta u$ & -$\frac{1}{36}\tilde{a}_V+\frac{1}{4}\tilde{a}_S $ &
330 & 950 & 750 \\ \cline{2-8}
 (uds)& $~~d~~$ & $\frac{1}{12}a_V+\frac{1}{4}a_S $ &
 $\Delta d$ & -$\frac{1}{36}\tilde{a}_V+\frac{1}{4}\tilde{a}_S $ &
330 & 950 & 750 \\ \cline{2-8}
 $~~~~$ & $~~s~~$ & $\frac{1}{3}a_V$ &
 $\Delta s$ & -$\frac{1}{9}\tilde{a}_V $ &
480 & 800 & 600 \\ \cline{1-8} $~~~~\Sigma^{-}~~~~$ & $~~d~~$
&$\frac{1}{6}a_V+\frac{1}{2}a_S $ &
 $\Delta d$ & -$\frac{1}{18}\tilde{a}_V+\frac{1}{2}\tilde{a}_S $ &
330 & 950 & 750 \\ \cline{2-8}
 (dds)& $~~s~~$ &$\frac{1}{3}a_V$ &
 $\Delta s$ & -$\frac{1}{9}\tilde{a}_V$ &
480 & 800 & 600 \\ \cline{1-8} $~~~~\Lambda^{0}~~~~$ & $~~u~~$
&$\frac{1}{4}a_V+\frac{1}{12}a_S $ &
 $\Delta u$ & -$\frac{1}{12}\tilde{a}_V+\frac{1}{12}\tilde{a}_S $ &
330 & 950 & 750 \\ \cline{2-8}
 (uds)& $~~d~~$ & $\frac{1}{4}a_V+\frac{1}{12}a_S $ &
 $\Delta d$ & -$\frac{1}{12}\tilde{a}_V+\frac{1}{12}\tilde{a}_S $ &
330 & 950 & 750 \\ \cline{2-8}
 $~~~~$ & $~~s~~$ & $\frac{1}{3}a_S$ &
 $\Delta s$ & $\frac{1}{3}\tilde{a}_S $ &
480 & 800 & 600 \\ \cline{1-8}
 $~~~~\Xi^{-}~~~~$ & $~~d~~$ &$\frac{1}{3}a_V $ &
 $\Delta d$ & -$\frac{1}{9}\tilde{a}_V $ &
330 & 1100 & 900 \\ \cline{2-8}
 (dss)& $~~s~~$ &$\frac{1}{6}a_V+\frac{1}{2} a_S$ &
 $\Delta s$ & -$\frac{1}{18}\tilde{a}_V+\frac{1}{2}\tilde{a}_S$ &
480 & 950 & 750 \\ \cline{1-8}
 $~~~~\Xi^{0}~~~~$ & $~~u~~$ &$\frac{1}{3}a_V $ &
 $\Delta u$ & -$\frac{1}{9}\tilde{a}_V $ &
330 & 1100 & 900 \\ \cline{2-8}
 (uss)& $~~s~~$ &$\frac{1}{6}a_V+\frac{1}{2} a_S$ &
 $\Delta s$ & -$\frac{1}{18}\tilde{a}_V+\frac{1}{2}\tilde{a}_S$ &
480 & 950 & 750 \\ \cline{1-8}
\end{tabular}
\end{center}
\end{footnotesize}

\vspace{0.5cm}

The quark distributions in baryons of the octet which  can be
calculated by using the parameters in Tab.~1, are  shown in
Figs.~1-8 ( $q_1$ and  $q_2$  for all octet baryons are specified
in Tab.~2 ). We need to mention that our results are consistent
with a recent calculation of the quark structure of the octet
baryons in the MIT bag model \cite{BT99}, though there are some
differences in the detailed features.



\subsection{Perturbative QCD Method}

We now extend the pQCD analysis of the quark structure
of baryons from the $\Lambda$ case \cite{MSY3} to all of the octet
baryons.
We adopt the canonical form for the quark distributions, following
Ref.~\cite{Bro95},
\begin{equation}
\begin{array}{cllr}
&q^{\uparrow}_{i}(x)=\frac{\tilde{A}_{q_{i}}}{B_3}
x^{-\frac{1}{2}}(1-x)^3+\frac{\tilde{B}_{q_{i}}}{B_4}
x^{-\frac{1}{2}}(1-x)^4;\\
&q^{\downarrow}_{i}(x)=\frac{\tilde{C}_{q_{i}}}{B_5}
x^{-\frac{1}{2}}(1-x)^5+\frac{\tilde{D}_{q_{i}}}{B_6}
x^{-\frac{1}{2}}(1-x)^6.
\end{array}
\label{case3}
\end{equation}
with  $i=1, 2$, where  $B_n=B(1/2,n+1)$ is the $\beta$-function
defined by $B(1-\alpha,n+1)=\int_0^1 x^{-\alpha}(1-x)^{n} {\mathrm
d} x$ for $\alpha=1/2$. From (\ref{case3}),  we obtain  the
valence quark normalization for  quark $q_i$
\begin{equation}
\begin{array}{clllc}
N_i=\tilde{A}_{q_{i}} +\tilde{B}_{q_{i}} +\tilde{C}_{q_{i}}
+\tilde{D}_{q_{i}},
\end{array}
\label{Ni}
\end{equation}
and the corresponding polarized distribution in the
$J^p=\frac{1}{2}^+$ octet
\begin{equation}
\begin{array}{clllc}
\Delta Q_{i}&=&\tilde{A}_{q_{i}} +\tilde{B}_{q_{i}}
-\tilde{C}_{q_{i}}  -\tilde{D}_{q_{i}},
\end{array}
\label{DQi}
\end{equation}
which can be extracted by using SU(3) symmetry from $\Sigma
Q_i=\Delta u +\Delta d +\Delta s \approx 0.20$ obtained in
deep-inelastic lepton-proton scattering experiments \cite{SMC95}
and the constants $F=0.459$ and $D=0.798$ obtained from hyperon
decay experiments \cite{Barnett96}.

In a strict sense, the $\Delta Q_{i}$ obtained this way from SU(3)
symmetry of the octet baryons should include both the
contributions from the valence and sea quarks. However, we know
from the recent measurements from semi-inclusive charged meson
production in DIS process \cite{SMC96,HERMES99} that the sea polarizations
are consistent with zero, and the measured $\Delta u$ and $\Delta
d$ for the valence quarks are close to the $\Delta u$ and $\Delta
d$ above. We may simply adopt the pQCD case 2 of Ref.\cite{MSY3}
with only the leading term for valence quarks (i.e., we set
$B_i=0$ and $D_i=0$) as an example for the pQCD predictions of the
quark distributions. However, to reflect the situation
that the real quark distributions
are actually more complicated, we shall adopt the pQCD case 3 with the
quark helicity sums $\Delta Q_1$ and $\Delta Q_2$ as
given in Tab.~2 to parameterize the quark distributions, based on
the above forms Eq.~(\ref{case3}) for the valence quark distributions.
In Tab.~2, the ratio
\begin{equation}
R_A=\frac{\tilde{A}_{q_{2}}}{\tilde{A}_{q_{1}}}\label{RA}
\end{equation}
reflects the $x \to 1 $ behaviour of
$\frac{q^{\uparrow}_{2}}{q^{\uparrow}_{1}}(x)$ in a baryon. For
every baryon, there are five constraints given by
Eqs.(\ref{Ni})-(\ref{RA}), which then result in three free
parameters, chosen as $\tilde{A}_{q_{1}}$, $\tilde{C}_{q_{1}}$ and
$\tilde{C}_{q_{2}}$, which should be actually further constrained
by relevant data. Here we are not intend to determine these
parameters with their exact values. However, we find, as an
example, one set of  parameters with the values
$\tilde{A}_{q_{1}}=5$, $\tilde{C}_{q_{1}}=3$ and
$\tilde{C}_{q_{2}}=2$ for the nucleon can give a rough shape of
quark flavor and spin distributions in the nucleon. We also find
that the parameters with the values $\tilde{A}_{q_{1}}=2$,
$\tilde{C}_{q_{1}}=2$ and $\tilde{C}_{q_{2}}=2$ for the $\Lambda$
can be used to give a good description of the
$\Lambda$-polarization \cite{MSY3}. Since we know little about the
quark distributions in the $\Sigma^0$, its input parameters are
taken to be the same as those for the $\Lambda^0$. The input
parameters for the other octet baryons are taken  to be the same
as the nucleon in consideration of the symmetry among them. All
other parameters determined according to the constraint conditions
are listed in Tab.~2.

\vspace{0.5cm}

\centerline{Table 2~~ The  parameters for quark distributions of
octet baryons  in pQCD}

\vspace{0.3cm}

\begin{footnotesize}
\begin{center}
\begin{tabular}{|c|c|c|c||c|c|c|c|c|c|c|c|c|c|}\hline
 Baryon & $q_1$ & $q_2$ & $R_A$ & $\Delta Q_1$ &
 $\Delta Q_2$ & $\tilde{A}_{q_1}$ & $\tilde{B}_{q_1}$
 &$\tilde{C}_{q_1}$ &$\tilde{D}_{q_1}$ &$\tilde{A}_{q_2}$
 &$\tilde{B}_{q_2}$ &$\tilde{C}_{q_2}$ &$\tilde{D}_{q_2}$\\ \hline
p & u & d & $\frac{1}{5}$ & 0.79 & -0.47 & 5.0 & -3.61 & 3.0 &
-2.40 & 1.0 & -0.74 & 2.0 & -1.27  \\ \hline n & d & u &
$\frac{1}{5}$ & 0.79 & -0.47 & 5.0 & -3.61 & 3.0 & -2.40 & 1.0 &
-0.74 & 2.0 & -1.27
\\ \hline $\Sigma^{+}$ & u & s & $\frac{1}{5}$ & 0.79 & -0.47 & 5.0 & -3.61 & 3.0 &
-2.40 & 1.0 & -0.74 & 2.0 & -1.27  \\ \hline $\Sigma^{0}$ & u(d) &
s & $\frac{2}{5}$ & 0.33 & -0.47 & 2.0 & -1.34 & 2.0 & -1.67 & 0.8
& -0.54 & 2.0 & -1.27
\\ \hline $\Sigma^{-}$ & d  & s & $\frac{1}{5}$ & 0.79 & -0.47 & 5.0 & -3.61
& 3.0 & -2.40 & 1.0 & -0.74 & 2.0 & -1.27  \\ \hline $\Lambda^{0}$
& s &  u(d) &  $\frac{1}{2}$  &0.60 &-0.20 & 2.0 & -1.20 & 2.0 &
-1.80 & 1.0 & -0.60 & 2.0 & -1.40  \\ \hline $\Xi^{-}$ & s & d
&$\frac{1}{5}$ & 0.79 & -0.47 & 5.0 & -3.61 & 3.0 & -2.40 & 1.0 &
-0.74 & 2.0 & -1.27
\\ \hline
$\Xi^{0}$ & s & u & $\frac{1}{5}$ & 0.79 & -0.47 & 5.0 & -3.61 &
3.0 & -2.40 & 1.0 & -0.74 & 2.0 & -1.27  \\ \hline
\end{tabular}
\end{center}
\end{footnotesize}

\vspace{0.5cm}

The quark distributions in octet baryons  calculated according to
the parameters in Tab.~2 are shown in Figs.~1-8. By comparing the
results (see Figs.1-8) obtained by using the SU(6) quark-diquark
model and the pQCD based model,  one can find the following
interesting features at $x \to 1$, for all of the other octet
baryons besides $\Lambda^0$: $$
   \left\{
      \begin{array}{ll}
      (\frac{\Delta q_2(x)}{q_2(x)})_{Diquark}
      & \rightarrow -\frac{1}{3};\\
      (\frac{\Delta q_2(x)}{q_2(x)})_{pQCD}
      & \rightarrow ~~1.
      \end{array}
\right. $$
 In addition, the two models also predict  different quark flavor
structures
$$
   \left\{
      \begin{array}{ll}
      (\frac{q_2(x)}{q_1(x)})_{Diquark}
      & \rightarrow 0;\\
      (\frac{q_2(x)}{q_1(x)})_{pQCD}
      & \rightarrow \frac{1}{5},
      \end{array}
\right. $$
for all other octet baryons besides $\Sigma^0$ and $\Lambda$.
This corresponds to  $s(x)/u(x)$ in $\Sigma^{\pm}$.
In case of the $\Sigma^0$ with $q_1=u, d$ and $q_2=s$, we find
$$
   \left\{
      \begin{array}{ll}
      (\frac{q_2(x)}{q_1(x)})_{Diquark}
      & \rightarrow 0;\\
      (\frac{q_2(x)}{q_1(x)})_{pQCD}
      & \rightarrow \frac{2}{5},
      \end{array}
\right. $$
which is with bigger difference of the flavor structure at large $x$
between the two models.
However, for the $\Lambda$ with $q_1=s$ and $q_2=u, d$,
we know that \cite{MSY2,MSY3} $$
   \left\{
      \begin{array}{ll}
      (\frac{q_2(x)}{q_1(x)})_{Diquark}
      & \rightarrow 0;\\
      (\frac{q_2(x)}{q_1(x)})_{pQCD}
      & \rightarrow \frac{1}{2},
      \end{array}
\right. $$ which has the largest difference in the flavor
structure at large $x$ between the two models among all the
baryons. This supports the conclusion in Ref.\cite{MSY2} that the
spin and flavor structure for the $\Lambda$ can provide a clean
test of different models. We notice that $q_1$ in Tab.~2 in both
models is the dominant quark contributing to the main spin and
valence structure at large $x$, and also that: $$
   \left\{
      \begin{array}{ll}
      (\frac{\Delta q_1(x)}{q_1(x)})_{Diquark}
      & \rightarrow 1;\\
      (\frac{\Delta q_1(x)}{q_1(x)})_{pQCD}
      & \rightarrow 1.
      \end{array}
\right. $$ Therefore the difference between the two models mainly
come from the valence quark $q_2$. We have also
neglected  the contribution of sea quarks and their polarizations,
although they may have  small contribution at small $x$.

It is
interesting to notice that the $\Sigma$'s have the most
significant difference in the flavor and spin structure between
the two models at medium to large $x$ region and this feature makes it
possible to have tests between pQCD and the quark-diquark model
predictions, as we will discuss in the following section.
It is even more appealing that the $\Lambda$ and $\Sigma^0$ have
complete different flavor and spin structure (remember
also that $q_1=u,d$ and $q_2=s$ for the $\Sigma^0$ whereas
$q_1=s$ and $q_2=u,d$ for the $\Lambda$) though they are
composed of same flavor quarks. Thus it is more novel to check the
different predictions concerning the flavor and spin structure
of the $\Lambda$ and $\Sigma^0$, or $\Sigma^{\pm}$. The
$\Sigma^\pm$ has close flavor structure compared to that
of $\Sigma^0$ (even we chose different sets of input parameters
as can be found from Tab.~2),
thus measurement of $\Sigma^\pm$ is also
helpful for our understanding of the $\Sigma^0$ structure.

\section{Drell-Yan Process for $\Sigma^{\pm}$ Beams on Isoscalar
Targets}

Although the $\Lambda$ can provide a clean test of the different
flavor and spin structure between the two different models, it is
still not possible to make a clear distinction between the above
two predictions with the available data of the
$\Lambda$-polarization in  the $e^+e^-$ process near the $Z$-pole
\cite{MSY3}. We also need a connection between the quark
distributions inside the $\Lambda$ and the  quark fragmentation
into a $\Lambda$ and such a connection is still not free from
theoretical and experimental uncertainties. As we pointed out in
the introduction, direct measurement of the quark structure of a
charged octet baryon is free from the requirement of a
connection between the quark distributions and the quark
fragmentation functions, in order to study the quark
structure of other baryons other than the nucleon.
It was also pointed out in
Ref.\cite{MSY2} that the charged particles, $\Sigma^{\pm}$ or
$\Xi^-$, may be used as beam in the Drell-Yan process, to test
different predictions concerning the quark structure of the
involved baryons. The Drell-Yan process has been widely used
experimentally to study the quark structure of the nucleon
\cite{McG99}. Using the $\Sigma^{\pm}$ as beam in Drell-Yan
processes  has been suggested \cite{Alberg96} for the
purpose of studying  the flavor asymmetry in the sea of the
baryons. In this section, we consider the Drell-Yan process  for
$\Sigma^{\pm}$ on isoscalar targets and show that it is also a
useful tool in order to study the flavor structure of
$\Sigma^{\pm}$ at medium to large $x$, in order to test
different predictions of the pQCD model and the quark spectator
diquark model.

For the process
\begin{equation}
\Sigma N \to l^+ l^- X,
\end{equation}
the cross section can be written as
\begin{equation}
\sigma (\Sigma N)=\frac{8 \pi \alpha^2}{9 \sqrt{\tau}} K(x_1,x_2)
\sum\limits_f e_f^2 [q_f^{\Sigma}
(x_1)\bar{q}_f(x_2)+\bar{q}_f^{\Sigma}(x_1) q_f(x_2)],
\end{equation}
where $\sqrt{\tau}=M/\sqrt{s}$, $M$ is the mass of the dilepton
pair. The factor $K(x_1,x_2)$ is due to higher-order QCD
corrections. More specifically,
\begin{eqnarray}
\sigma(\Sigma^+ p)&=&\frac{8 \pi \alpha^2}{9 \sqrt{\tau}}
K(x_1,x_2) \{ \frac{4}{9} [u^{\Sigma^+}(x_1)\bar{u}(x_2)+
\bar{u}^{\Sigma^+}(x_1)u(x_2)] \nonumber\\
&+
&\frac{1}{9}[d^{\Sigma^+}(x_1) \bar{d}_(x_2)+
\bar{d}^{\Sigma^+}(x_1) d(x_2)]
\nonumber\\
&+&\frac{1}{9}[s^{\Sigma^+}(x_1) \bar{s}_(x_2)+
\bar{s}^{\Sigma^+}(x_1) s(x_2) ]\},
\end{eqnarray}
and
\begin{eqnarray}
\sigma(\Sigma^+ n) &=& \frac{8 \pi \alpha^2}{9 \sqrt{\tau}}
K(x_1,x_2) \{ \frac{4}{9} [u^{\Sigma^+}(x_1)\bar{d}(x_2)+
\bar{u}^{\Sigma^+}(x_1)d(x_2)] \nonumber \\
&+&\frac{1}{9}[d^{\Sigma^+}(x_1) \bar{u}_(x_2)+
\bar{d}^{\Sigma^+}(x_1) u(x_2) ]
\nonumber\\
&+&\frac{1}{9}[s^{\Sigma^+}(x_1) \bar{s}_(x_2)+
\bar{s}^{\Sigma^+}(x_1) s(x_2) ]\}.
\end{eqnarray}

By using charge symmetry, i.e.,
$u^{\Sigma^+} \leftrightarrow d^{\Sigma^-}$,
$d^{\Sigma^+} \leftrightarrow u^{\Sigma^-}$,
$\bar{u}^{\Sigma^+} \leftrightarrow \bar{d}^{\Sigma^-}$,
and $\bar{d}^{\Sigma^+} \leftrightarrow \bar{u}^{\Sigma^-}$,
one can obtain the cross section of
$\Sigma^- N$ as
\begin{eqnarray}
\sigma(\Sigma^- p)&=&\frac{8 \pi \alpha^2}{9 \sqrt{\tau}}
K(x_1,x_2) \{ \frac{1}{9} [u^{\Sigma^+}(x_1)\bar{d}(x_2)+
\bar{u}^{\Sigma^+}(x_1)d(x_2)]
\nonumber\\
&+&\frac{4}{9}[d^{\Sigma^+}(x_1) \bar{u}_(x_2)+
\bar{d}^{\Sigma^+}(x_1) u(x_2) ]
\nonumber\\
&+&\frac{1}{9}[s^{\Sigma^+}(x_1)\bar{s}(x_2)+
\bar{s}^{\Sigma^+}(x_1) s(x_2) ]
\},
\end{eqnarray}
and
\begin{eqnarray}
\sigma(\Sigma^- n)&=&\frac{8 \pi \alpha^2}{9 \sqrt{\tau}}
K(x_1,x_2) \{ \frac{1}{9} [u^{\Sigma^+}(x_1)\bar{u}(x_2)+
\bar{u}^{\Sigma^+}(x_1)u(x_2)]
\nonumber\\
&+&\frac{4}{9}[d^{\Sigma^+}(x_1)\bar{d}(x_2)+
\bar{d}^{\Sigma^+}(x_1) d(x_2) ]
\nonumber\\
&+&\frac{1}{9}[s^{\Sigma^+}(x_1)\bar{s}(x_2)+
\bar{s}^{\Sigma^+}(x_1) s(x_2) ]
\}.
\end{eqnarray}

Considering  an isoscalar target with the nucleon number of $A$, we
obtain the cross section
\begin{eqnarray}
\sigma^+= \sigma(\Sigma^+ A)&=&\frac{8 \pi \alpha^2}{9
\sqrt{\tau}} \frac{A}{2} K(x_1,x_2) \{ \frac{4}{9}
[u^{\Sigma^+}(x_1)\bar{u}(x_2)+
\bar{u}^{\Sigma^+}(x_1)u(x_2)\nonumber \\ &+&
u^{\Sigma^+}(x_1)\bar{d}(x_2)+ \bar{u}^{\Sigma^+}(x_1)d(x_2)]
\nonumber \\ &+&\frac{1}{9}[d^{\Sigma^+}(x_1) \bar{d}_(x_2)
+d^{\Sigma^+}(x_1) \bar{u}_(x_2)
+\bar{d}^{\Sigma^+}(x_1) d(x_2)
+ \bar{d}^{\Sigma^+}(x_1) u(x_2)]
\nonumber \\ &+&\frac{1}{9}[2 s^{\Sigma^+}(x_1) \bar{s}_(x_2)+
2 \bar{s}^{\Sigma^+}(x_1) s_(x_2)]
\} ;
\label{S1}
\end{eqnarray}
\begin{eqnarray}
\sigma^-=\sigma(\Sigma^- A)&=&\frac{8 \pi \alpha^2}{9 \sqrt{\tau}}
\frac{A}{2} K(x_1,x_2) \{ \frac{1}{9}
[u^{\Sigma^+}(x_1)\bar{d}(x_2)+
\bar{u}^{\Sigma^+}(x_1)d(x_2)\nonumber\\&+&u^{\Sigma^+}(x_1)\bar{u}(x_2)
+\bar{u}^{\Sigma^+}(x_1)u(x_2)]
\nonumber \\
&+&\frac{4}{9}[d^{\Sigma^+}(x_1) \bar{u}_(x_2)+
d^{\Sigma^+}(x_1) \bar{d}_(x_2)+\bar{d}^{\Sigma^+}(x_1)
u(x_2)+\bar{d}^{\Sigma^+}(x_1) d(x_2) ]
\nonumber \\ &+&\frac{4}{9}[2 s^{\Sigma^+}(x_1) \bar{s}_(x_2)+
2 \bar{s}^{\Sigma^+}(x_1) s_(x_2)]
\} .
\label{S2}
\end{eqnarray}

Choice of different ranges of the variables $x_1$ and $x_2$ and
different combination of the targets with different isospin
properties can help us to pin down the information for the various
quark distributions of the $\Sigma^{\pm}$. The purpose of this
paper is to study the valence quark structure of the
$\Sigma^{\pm}$ at medium to large $x$. Considering the fact that
valence quarks dominate in the hyperons at medium to large $x$, we
obtain
\begin{equation}
\tilde{\sigma}^+(x_1,x_2)=\frac{8 \pi \alpha^2}{9 \sqrt{\tau}}
\frac{A}{2} K(x_1,x_2) \{ \frac{4}{9}
[u^{\Sigma^+}(x_1)[\bar{u}(x_2)+ \bar{d}(x_2)]+ \frac{2}{9}
s^{\Sigma^+}(x_1) \bar{s}(x_2)\}; \label{ep1}
\end{equation}
\begin{equation}
\tilde{\sigma}^-(x_1,x_2)=\frac{8 \pi \alpha^2}{9 \sqrt{\tau}}
\frac{A}{2} K(x_1,x_2) \{ \frac{1}{9}
u^{\Sigma^+}(x_1)[\bar{u}(x_2)+\bar{d}(x_2)]+ \frac{2}{9}
s^{\Sigma^+}(x_1)\bar{s}(x_2)\}. \label{ep2}
\end{equation}
Furthermore, we
introduce the ratio
\begin{eqnarray}
T_V(x_1,x_2)&=&\frac{\tilde{\sigma}^+(x_1,x_2)}{\tilde{\sigma}^-(x_1,x_2)}
=\displaystyle{\frac{4[\frac{u(x_1)}{s(x_1)}]^{\Sigma^+}+\kappa(x_2)}
{[\frac{u(x_1)}{s(x_1)}]^{\Sigma^+}+\kappa(x_2)}},
\end{eqnarray}
where the ratio
$\kappa(x_2)=2\bar{s}(x_2)/(\bar{u}(x_2)+\bar{d}(x_2))$, which
denotes the strange quark content of the nucleon.
$\kappa=2\bar{s}/(\bar{u}+\bar{d})$ with $x_2$ in the quark
distributions being integrated has been determined experimentally
in neutrino-induced charm production\cite{Abrom82,Fouds90,Stron91}
to be in the range $0.373^{+0.048}_{-0.041}\pm 0.018 \leq \kappa
\leq 0.57 \pm 0.09$. Actually, we find that the ratio
$\kappa(x_2)$ varies from 0.547 to 0.404 while  $x_2$ increasing
from 0.001 to 0.7 according to the parameterizations of CTEQ
\cite{CTEQ}. We take $x_2=0.3$ and then the value of $\kappa(x_2)$
is  $0.44$, the predictions of $T_V$  both in pQCD based and SU(6)
diquark spectator models are shown in Fig.~9. The deviation
between pQCD and diquark predictions for $T_V$ is large enough in
order to be used to test the flavor structure of $\Sigma^\pm$. We
point out here that the above $T_V(x_1,x_2)$ is only one example
for the quantities that could show the difference between the two
predictions, other quantities, such as
$\frac{{\sigma}(\Sigma^+p)}{{\sigma}(\Sigma^+n)}$, could be an
alternative quantity to test different predictions.

The use of the variables $x_1$ and $x_2$ is convenient for
theoretical calculations and they signify the light-cone momentum
fractions carried by the two colliding quarks in the beam hadron
and target hadron. In the experimental measurements, it is
convenient to use the experimental variables $\tau=M^2/s$, where
$M$ is the mass of the dilepton pair and $s$ is the square of the
total energy in the center of mass frame, and $y$, which is the
rapidity of the dilepton pair, instead of the variables $x_1$ and
$x_2$. The two variables $\tau$ and $y$, when expressed in terms
of $x_1$ and $x_2$, read
\begin{eqnarray}
&&\tau=x_1 x_2; \nonumber\\ &&y=\frac{1}{2} \ln(\frac{x_1}{x_2}),
\end{eqnarray}
from which we get the two variables $x_1$ and $x_2$ in terms of
$\tau$ and $y$
\begin{eqnarray}
&&x_1=\rm{e}^y \sqrt{\tau}; \nonumber \\
&&x_2=\rm{e}^{-y}\sqrt{\tau}.
\end{eqnarray}
Thus we can express the quantity $T_V$ as a function of $\tau$ and
$y$. It will be helpful if we can measure $T_V(\tau,y)$ near
the experimentally most accessible values of $\tau$ and $y$, and
also find significant differences between the predictions of  the
two  models. We present $T_V(\tau,y)$ in Figs.~10-11 for  two
cases: (i) at fixed $y=0$ by using $\tau$ as a variable in
Fig.~10, and (ii) at fixed $\tau=0.02$ by using $y$ as a variable
in Fig.~11. We find significant differences between the two
predictions, thus it is possible to check the $\Sigma^{\pm}$
flavor structure by measuring the quantity $T_V(\tau,y)$.

In principle we may extend the discussion to the case where the
charged octet baryon $\Xi^-$ is used as the beam. In this case $s$
is the dominant valence quark and $d$ is the less dominant valence
quark at large $x$ inside $\Xi^-$. Thus for the Drell-Yan process
\begin{equation}
\Xi^{-} N \to l^+ l^- X,
\end{equation}
the cross sections can be written as,
\begin{eqnarray}
\sigma(\Xi^- p)&=&\frac{8 \pi \alpha^2}{9 \sqrt{\tau}}
K(x_1,x_2) \{ \frac{4}{9} [u^{\Xi^-}(x_1)\bar{u}(x_2)+
\bar{u}^{\Xi^-}(x_1)u(x_2)] \nonumber\\
&+
&\frac{1}{9}[d^{\Xi^-}(x_1) \bar{d}_(x_2)+
\bar{d}^{\Xi^-}(x_1) d(x_2)]
\nonumber\\
&+&\frac{1}{9}[s^{\Xi^-}(x_1) \bar{s}_(x_2)+
\bar{s}^{\Xi^-}(x_1) s(x_2) ]\},
\label{xi1}
\end{eqnarray}
and
\begin{eqnarray}
\sigma(\Xi^- n) &=& \frac{8 \pi \alpha^2}{9 \sqrt{\tau}}
K(x_1,x_2) \{ \frac{4}{9} [u^{\Xi^-}(x_1)\bar{d}(x_2)+
\bar{u}^{\Xi^-}(x_1)d(x_2)] \nonumber \\
&+&\frac{1}{9}[d^{\Xi^-}(x_1) \bar{u}_(x_2)+
\bar{d}^{\Xi^-}(x_1) u(x_2) ]
\nonumber\\
&+&\frac{1}{9}[s^{\Xi^-}(x_1) \bar{s}_(x_2)+
\bar{s}^{\Xi^-}(x_1) s(x_2) ]\}.
\label{xi2}
\end{eqnarray}
Considering the fact that
the valence quarks $s$ and $d$ dominate in the $\Xi^-$ at large $x$,
we obtain
\begin{equation}
\tilde{\sigma}^p(x_1,x_2)=\frac{8 \pi \alpha^2}{9 \sqrt{\tau}}
K(x_1,x_2) \{ \frac{1}{9} d^{\Xi^-}(x_1)\bar{d}(x_2) +\frac{1}{9}
s^{\Xi^-}(x_1) \bar{s}(x_2) \}; \label{epxi1}
\end{equation}
\begin{equation}
\tilde{\sigma}^n(x_1,x_2)=\frac{8 \pi \alpha^2}{9 \sqrt{\tau}}
K(x_1,x_2) \{ \frac{1}{9} d^{\Xi^-}(x_1)\bar{u}(x_2) +\frac{1}{9}
s^{\Xi^-}(x_1)\bar{s}(x_2) \}, \label{epxi2}
\end{equation}
from which we have
\begin{equation}
\Delta \sigma=\tilde{\sigma}^p(x_1,x_2)-\tilde{\sigma}^n(x_1,x_2)
=\frac{8 \pi \alpha^2}{9 \sqrt{\tau}} K(x_1,x_2) \{ \frac{1}{9}
d^{\Xi^-}(x_1)[\bar{d}(x_2)-\bar{u}(x_2)] \}, \label{del}
\end{equation}
where the behavior of $\bar{d}(x_2)-\bar{u}(x_2)$ is relatively
well known from available studies on the Gottfried sum rule
violation \cite{Kum97}. Thus the valence quark distribution
$d^{\Xi^-}(x)$ can be measured by using Eq.~(\ref{del}), and then
substitute the measured $d^{\Xi^-}(x)$ into Eq.~(\ref{epxi1}) or
(\ref{epxi2}) one can obtain the dominant valence quark
$s^{\Xi^-}(x)$ and check different predictions, if the data
precision is high enough.

We would like to mention that in principle it should be possible
to measure the spin structure of $\Sigma^{\pm}$ from processes in
which polarized $\Sigma^{\pm}$ beams are involved, such as the
spin-dependent Drell-Yan process. However, it is difficult to
obtain polarized $\Sigma^{\pm}$ beams. Therefore, it might be
comparatively easier to use various $\Sigma^{\pm}$ fragmentation
processes as has been suggested for the $\Lambda$ case
\cite{MSY2,MSY3,Ma99} to study the spin and flavor structure of
$\Lambda^{\pm}$, though this may suffer from  uncertainties in the
connections between the quark distributions and fragmentation
functions. The spin structure of $\Sigma^{\pm}$  can be measured
from polarized $\Sigma^{\pm}$ fragmentations in processes such as
$e^+e^-$ annihilation near the $Z$-pole and semi-inclusive deep
inelastic scattering.

\section{Summary}

We found in this paper that $\Lambda$ has the most significant
difference in the flavor structure at large $x$, compared to that
of other octet baryons, between the SU(6) quark spectator diquark
model and the pQCD based model, and this supports the conclusion
in a previous study \cite{MSY2} that the $\Lambda$ can provide a
new domain to test different models concerning the flavor and spin
structure of the nucleon. However, for the $\Lambda$ we still need
a connection between the quark distributions inside the $\Lambda$
and a quark fragmentation into a $\Lambda$, due to the fact that
$\Lambda$ is a charge neutral and therefore cannot be used as beam
and its short lifetime makes it also difficult to serve as the
target. In order to avoid the theoretical and experimental
uncertainties concerning the connection between quark
distributions and quark fragmentations, it is meaningful to find
a charged baryon with also different flavor and spin structure
between the two models. In order to test the flavor structure of
octet baryons, we turn our attention in this paper on the flavor
structure of $\Sigma^{\pm}$, which has the most significant
difference between the two models at medium to large $x$. The
ratio $T_V$ of Drell-Yan total cross section of $\Sigma^+$ beam to
that of $\Sigma^-$ beam to the isoscalar targets is sensitive to
different models. The measurement of $T_V$ can offer further
information for a distinction between different predictions
concerning the flavor structure of octet baryons. It is more
interesting that the $\Lambda$ and $\Sigma^0$ have complete
different flavor and spin structure though they are composed of
same flavor quarks. The $\Sigma^\pm$ have similar  flavor
structure compared to that of $\Sigma^0$, thus measurement of
$\Sigma^\pm$ is also helpful for our understanding of the
$\Sigma^0$ structure.

{\bf Acknowledgments: }
This work is partially supported by
Fondecyt (Chile) postdoctoral fellowship 3990048, by Fondecyt
(Chile) grant 1990806 and by a C\'atedra Presidencial (Chile), and
by National Natural Science Foundation of China under Grant
Numbers 19605006, 19875024, 19775051, and 19975052.

\newpage

\begin{figure}[htb]
\begin{center}
\leavevmode {\epsfysize=5.5cm \epsffile{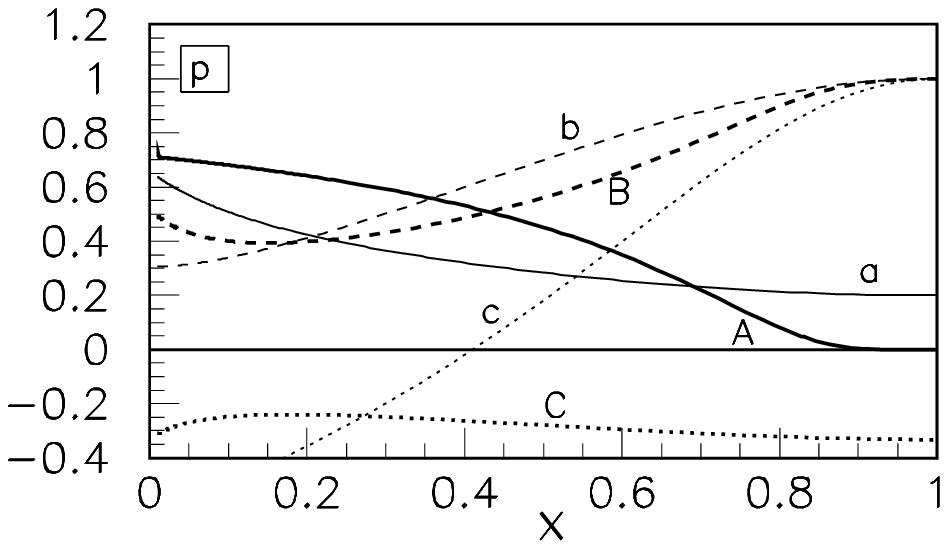}}
\end{center}
\caption[*]{\baselineskip 13pt $A=(\frac{ q_2}{q_1})^{Diquark}$,
$B=(\frac{\Delta q_1}{q_1})^{Diquark}$, $C=(\frac{\Delta
q_2}{q_2})^{Diquark}$,  $a=(\frac{ q_2}{q_1})^{pQCD}$,
$b=(\frac{\Delta q_1}{q_1})^{pQCD}$ and $c=(\frac{\Delta
q_2}{q_2})^{pQCD}$ for  the proton in which $q_1=u$ and $q_2=d$.
}\label{msy4f1}
\end{figure}

\begin{figure}[htb]
\begin{center}
\leavevmode {\epsfysize=5.5cm \epsffile{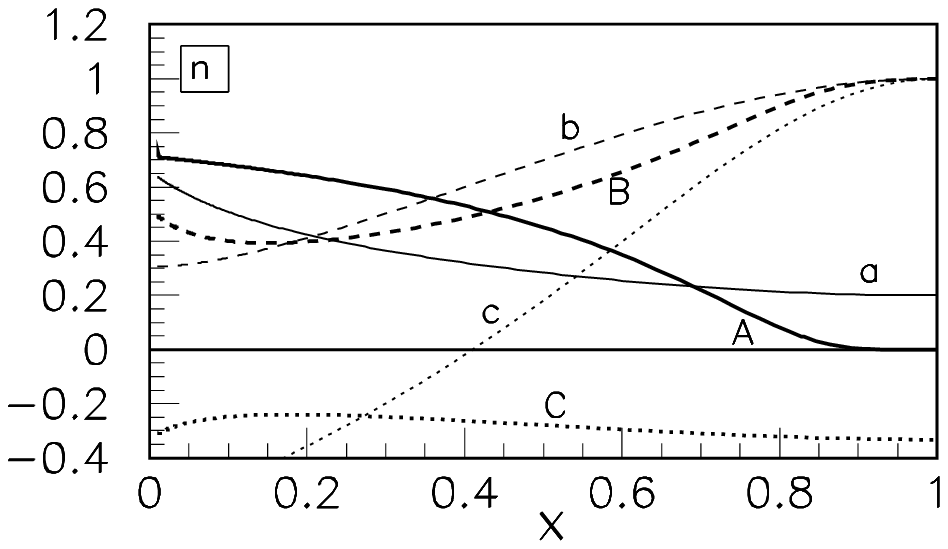}}
\end{center}
\caption[*]{\baselineskip 13pt $A=(\frac{ q_2}{q_1})^{Diquark}$,
$B=(\frac{\Delta q_1}{q_1})^{Diquark}$, $C=(\frac{\Delta
q_2}{q_2})^{Diquark}$, $a=(\frac{ q_2}{q_1})^{pQCD}$,
$b=(\frac{\Delta q_1}{q_1})^{pQCD}$ and $c=(\frac{\Delta
q_2}{q_2})^{pQCD}$ for  the neutron in which $q_1=d$ and $q_2=u$.
}\label{msy4f2}
\end{figure}

\begin{figure}[htb]
\begin{center}
\leavevmode {\epsfysize=5.5cm \epsffile{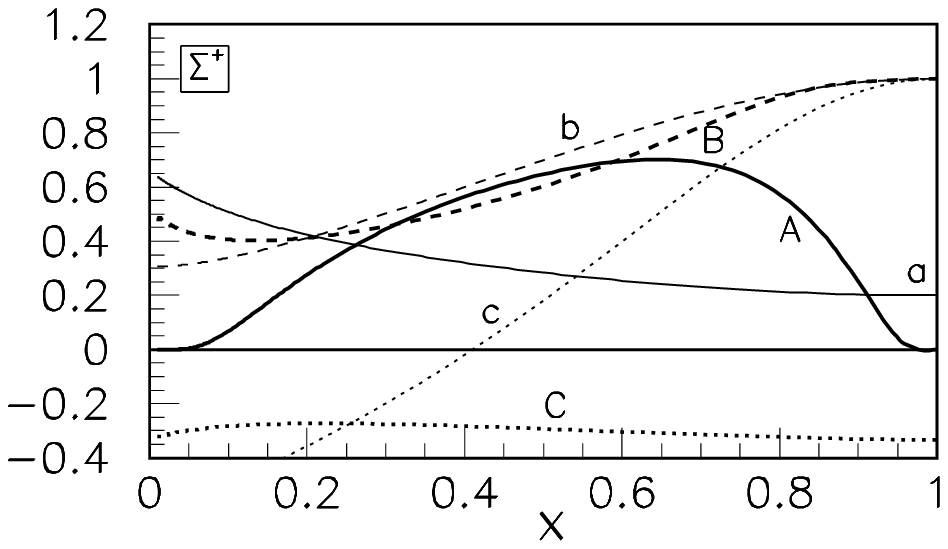}}
\end{center}
\caption[*]{\baselineskip 13pt $A=(\frac{ q_2}{q_1})^{Diquark}$,
$B=(\frac{\Delta q_1}{q_1})^{Diquark}$, $C=(\frac{\Delta
q_2}{q_2})^{Diquark}$, $a=(\frac{ q_2}{q_1})^{pQCD}$,
$b=(\frac{\Delta q_1}{q_1})^{pQCD}$ and $c=(\frac{\Delta
q_2}{q_2})^{pQCD}$ for  the $\Sigma^+$ in which $q_1=u$ and
$q_2=s$. }\label{msy4f3}
\end{figure}

\begin{figure}[htb]
\begin{center}
\leavevmode {\epsfysize=5.5cm \epsffile{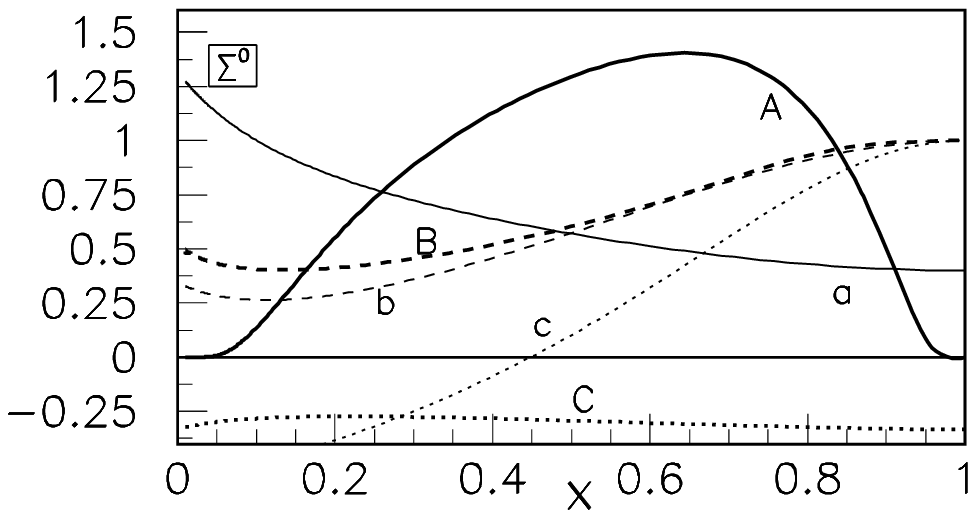}}
\end{center}
\caption[*]{\baselineskip 13pt $A=(\frac{ q_2}{q_1})^{Diquark}$,
$B=(\frac{\Delta q_1}{q_1})^{Diquark}$, $C=(\frac{\Delta
q_2}{q_2})^{Diquark}$, $a=(\frac{ q_2}{q_1})^{pQCD}$,
$b=(\frac{\Delta q_1}{q_1})^{pQCD}$ and $c=(\frac{\Delta
q_2}{q_2})^{pQCD}$ for  the $\Sigma^0$ in which $q_1=u(d)$ and
$q_2=s$. }\label{msy4f4}
\end{figure}

\begin{figure}[htb]
\begin{center}
\leavevmode {\epsfysize=5.5cm \epsffile{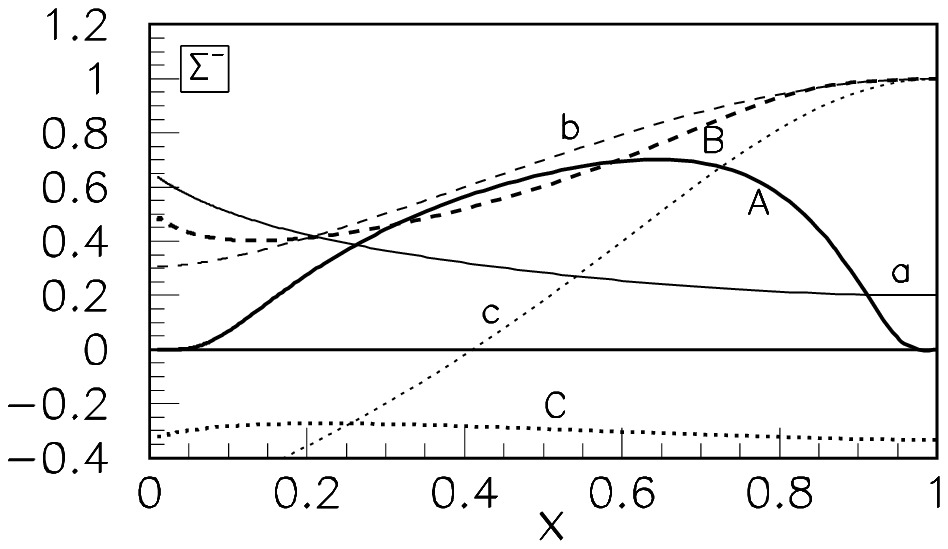}}
\end{center}
\caption[*]{\baselineskip 13pt $A=(\frac{ q_2}{q_1})^{Diquark}$,
$B=(\frac{\Delta q_1}{q_1})^{Diquark}$, $C=(\frac{\Delta
q_2}{q_2})^{Diquark}$, $a=(\frac{ q_2}{q_1})^{pQCD}$,
$b=(\frac{\Delta q_1}{q_1})^{pQCD}$ and $c=(\frac{\Delta
q_2}{q_2})^{pQCD}$ for  the $\Sigma^-$ in which $q_1=d$ and
$q_2=s$. }\label{msy4f5}
\end{figure}

\begin{figure}[htb]
\begin{center}
\leavevmode {\epsfysize=5.5cm \epsffile{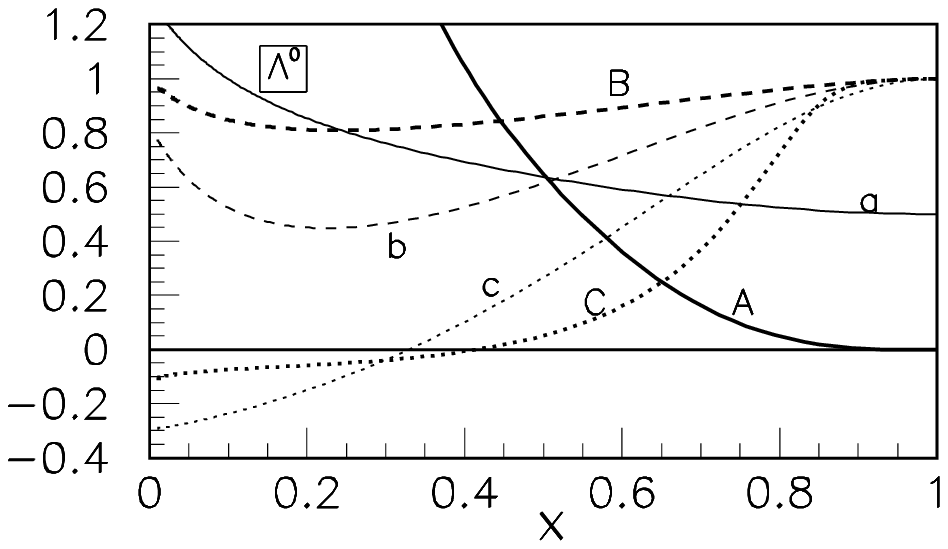}}
\end{center}
\caption[*]{\baselineskip 13pt $A=(\frac{ q_2}{q_1})^{Diquark}$,
$B=(\frac{\Delta q_1}{q_1})^{Diquark}$, $C=(\frac{\Delta
q_2}{q_2})^{Diquark}$, $a=(\frac{ q_2}{q_1})^{pQCD}$,
$b=(\frac{\Delta q_1}{q_1})^{pQCD}$ and $c=(\frac{\Delta
q_2}{q_2})^{pQCD}$ for  the $\Lambda^0$ in which $q_1=s$ and
$q_2=u(d)$. }\label{msy4f6}
\end{figure}

\begin{figure}[htb]
\begin{center}
\leavevmode {\epsfysize=5.5cm \epsffile{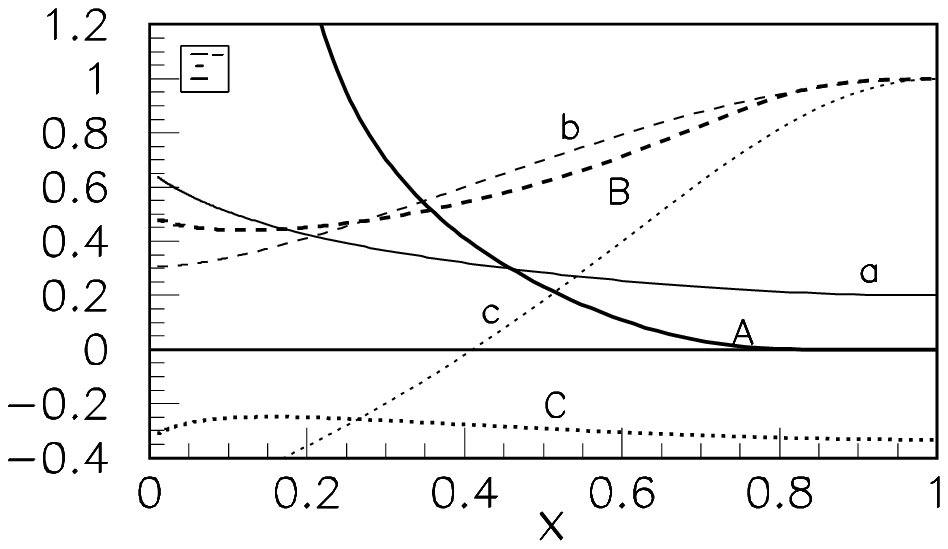}}
\end{center}
\caption[*]{\baselineskip 13pt $A=(\frac{ q_2}{q_1})^{Diquark}$,
$B=(\frac{\Delta q_1}{q_1})^{Diquark}$, $C=(\frac{\Delta
q_2}{q_2})^{Diquark}$, $a=(\frac{ q_2}{q_1})^{pQCD}$,
$b=(\frac{\Delta q_1}{q_1})^{pQCD}$ and $c=(\frac{\Delta
q_2}{q_2})^{pQCD}$ for  the $\Xi^-$ in which $q_1=s$ and $q_2=d$.
}\label{msy4f7}
\end{figure}

\begin{figure}[htb]
\begin{center}
\leavevmode {\epsfysize=5.5cm \epsffile{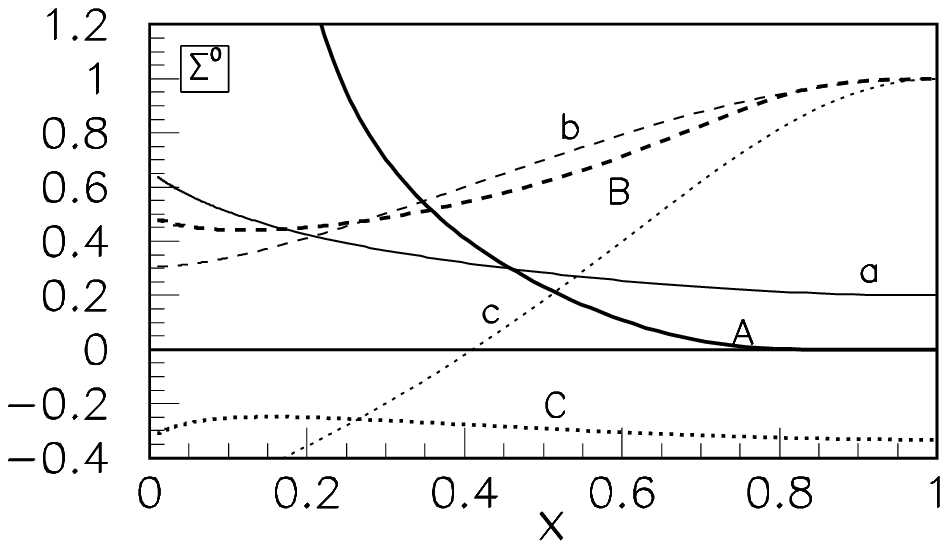}}
\end{center}
\caption[*]{\baselineskip 13pt $A=(\frac{ q_2}{q_1})^{Diquark}$,
$B=(\frac{\Delta q_1}{q_1})^{Diquark}$, $C=(\frac{\Delta
q_2}{q_2})^{Diquark}$, $a=(\frac{ q_2}{q_1})^{pQCD}$,
$b=(\frac{\Delta q_1}{q_1})^{pQCD}$ and $c=(\frac{\Delta
q_2}{q_2})^{pQCD}$ for  the $\Xi^0$ in which $q_1=s$ and $q_2=u$.
}\label{msy4f8}
\end{figure}

\begin{figure}[htb]
\begin{center}
\leavevmode {\epsfysize=5.5cm \epsffile{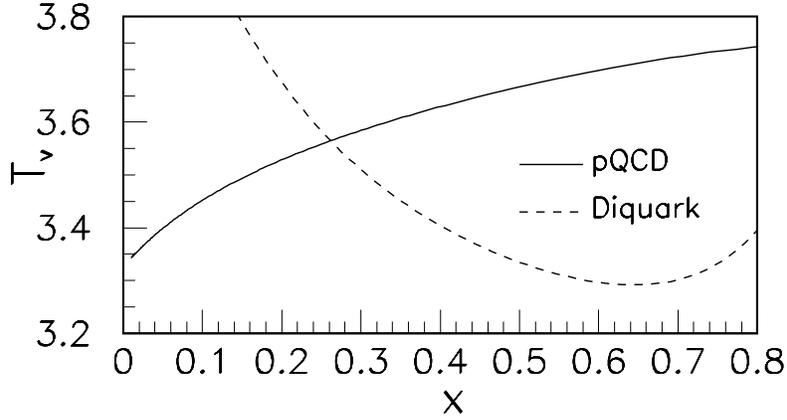}}
\end{center}
\caption[*]{\baselineskip 13pt The  $T_V(x_1,x_2)$ in the SU(6)
quark-diquark  and pQCD based Models at fixed $x_2=0.3$ as a
function of $x=x_1$. }\label{msy4f9}
\end{figure}

\begin{figure}[htb]
\begin{center}
\leavevmode {\epsfysize=5.5cm \epsffile{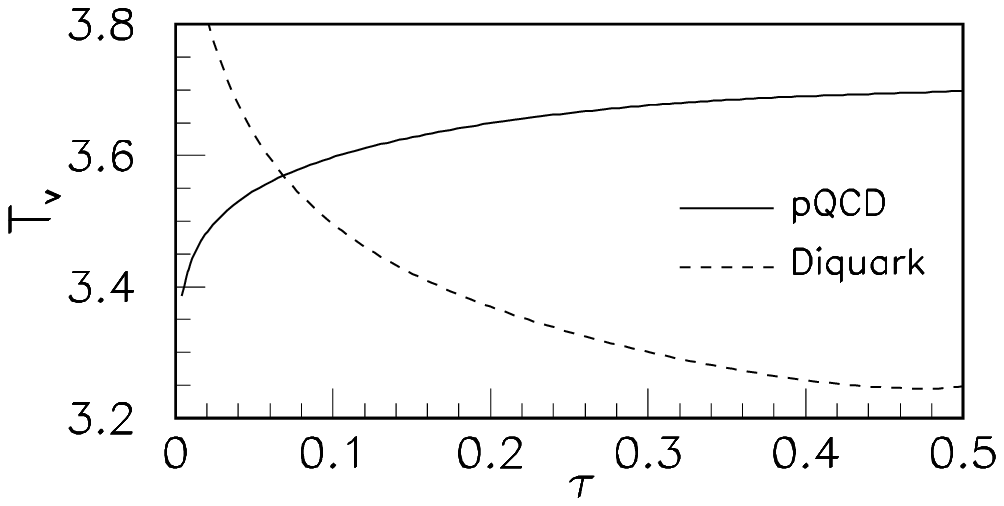}}
\end{center}
\caption[*]{\baselineskip 13pt The  $T_V(\tau,y)$ in the SU(6)
quark-diquark  and pQCD based Models at fixed $y=0$ as a function
of $\tau$. }\label{msy4f10}
\end{figure}

\begin{figure}[htb]
\begin{center}
\leavevmode {\epsfysize=5.5cm \epsffile{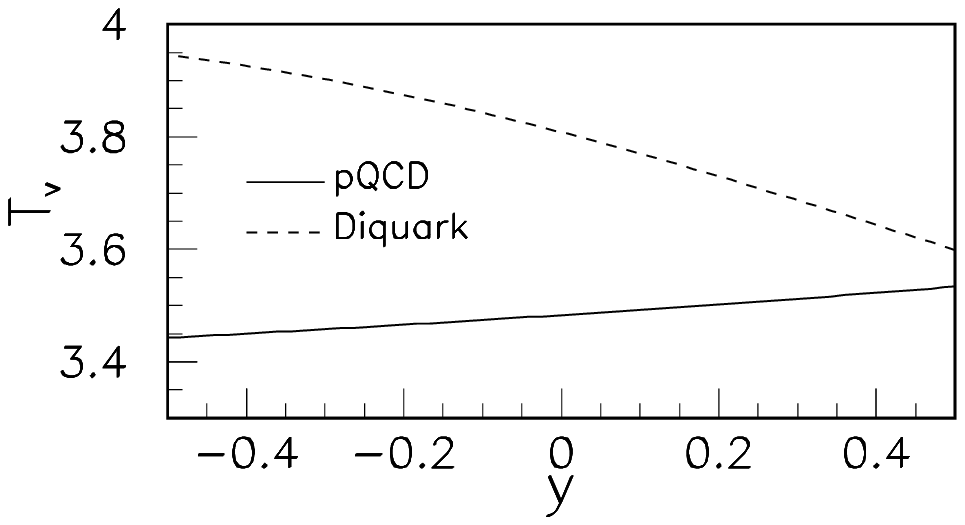}}
\end{center}
\caption[*]{\baselineskip 13pt The  $T_V(\tau,y)$ in the SU(6)
quark-diquark  and pQCD based Models at fixed $\tau=0.02$ as a
function of $y$. }\label{msy4f11}
\end{figure}


\newpage

\begin{thebibliography}{99}

\bibitem{SpinR}
For reviews, see, e.g., H.-Y. Cheng, Int.~J.~Mod.~Phys.~{\bf A
11} (1996) 5109; G.P. Ramsey, Prog.~Part.~Nucl.~Phys. {\bf 39}
(1997) 599. For a recent discussion, see, B.-Q.~Ma, I.~Schmidt,
and J.~Soffer, Phys. Lett. {\bf B 441} (1998) 461.

\bibitem{Bro88}
S.J.~Brodsky, J.~Ellis, and M.~Karliner,
Phys. Lett. {\bf B 206} (1988) 309;

J.~Ellis and M.~Karliner, Phys. Lett. {\bf B 213} (1988) 73; {\bf B
341} (1995) 397.

\bibitem{Bro96}
See, e.g., S.J.~Brodsky and B.-Q.~Ma, Phys.~Lett.~{\bf B 381} (1996) 317,
and references therein.

\bibitem{Kum97}
For a recent review, see, e.g., S. Kumano, Phys. Rep. {\bf 303}
 (1998) 183.

\bibitem{Isospin}
B.-Q. Ma, Phys. Lett. {\bf B 274} (1992) 111; C. Boros, J.T.
Londergan, and A.W. Thomas, Phys. Rev. Lett. {\bf 81}
(1998) 4075.

\bibitem{Far75}
G.R. Farrar and D.R. Jackson, Phys. Rev. Lett. {\bf 35}
(1975) 1416.

\bibitem{DQM}
R.~Carlitz, Phys.~Lett.~{\bf B 58} (1975) 345; J.~Kaur,
Nucl.~Phys.~{\bf B 128} (1977) 219; A.~Sch\"afer,
Phys.~Lett.~{\bf B 208} (1988) 175.

\bibitem{Bro95}
S.J. Brodsky, M. Burkardt, and I. Schmidt, Nucl. Phys. {\bf B
441} (1995) 197.

\bibitem{Ma96}
B.-Q. Ma, Phys. Lett. {\bf B 375} (1996) 320.

\bibitem{Mel96}
W. Melnitchouk and A.W. Thomas, Phys. Lett. {\bf B 377}
(1996) 11.

\bibitem{Yang99}
U.K. Yang and A. Bodek, Phys. Rev. Lett. {\bf 82} (1999) 2467.

\bibitem{MSY2}
B.-Q. Ma, I. Schmidt, and J.-J. Yang, 
Phys. Lett. {\bf B 477} (2000) 107.

\bibitem{MSY3}
B.-Q. Ma, I. Schmidt, and J.-J. Yang, 
Phys. Rev. {\bf D 61} (2000) 034017.

\bibitem{HERMES99b}
The HERMES Collab., A.~Airapetian {\it et al.}, 
hep-ex/9911017.

\bibitem{GLR}
V.N.~Gribov and L.N.~Lipatov, Phys. Lett. {\bf B 37} (1971) 78;
Sov. J. Nucl. Phys. {\bf 15} (1972) 675.

\bibitem{Bro97}
S.J.~Brodsky and B.-Q.~Ma, Phys. Lett. {\bf B 392} (1997) 452.

\bibitem{Ma99}
B.-Q. Ma and J. Soffer, Phys. Rev. Lett. {\bf 82} (1999) 2250.

\bibitem{Ma91b}
B.-Q.~Ma, J. Phys. {\bf G 17} (1991) L53;

B.-Q.~Ma and Q.-R.~Zhang, Z.~Phys. {\bf C 58} (1993) 479.

\bibitem{BHL}
S. J. Brodsky, T. Huang, and G. P. Lepage, in {\it Particles and
Fields-2}, Proceedings of the Banff Summer Institute, Banff,
Alberta, 1981, edited by A. Z. Capri and A. N. Kamal (Plenum, New
York,1983), p. 143.

\bibitem{BT99}
C. Boros and A. W. Thomas, Phys. Rev. {\bf D 60} (1999) 074017.


\bibitem{SMC95}
SM Collab., D. Adams et al., Phys. Lett. {\bf{B 357}} (1995) 248.

\bibitem{Barnett96} Particle Data Group, R. M. Barnett et al., Phys.
Rev. {\bf{D 54}} (1996) 1.

See, e.g.,
C.~Boros and Z.~Liang, Phys. Rev. {\bf D 57} (1998) 4491.


\bibitem{SMC96}
SM Collab., B. Adeva {\it et al.}, Phys. Lett. {\bf B 369}
(1996) 93; {\bf B 420} (1998) 180.

\bibitem{HERMES99}
The HERMES Collab., K. Ackerstaff et al.,
Phys. Lett. {\bf B 484} (1999) 123.

\bibitem{McG99}
For a recent review, see, e.g., P.L. McGaughey, J.M. Moss,
and J.C. Peng, 
Ann.Rev.Nucl.Part.Sci. {\bf 49} (1999) 217. 

\bibitem{Alberg96}
M. Alberg {\it et al},
Phys. Lett. {\bf{B 389}} (1996) 367.

\bibitem{Abrom82}
H. Abromowicz et al., Z. Phys. {\bf{C 15}} (1982) 19.

\bibitem{Fouds90}
C. Foudas et al. Phys. Rev. Lett. {\bf{64}} (1990) 1207; M.H.
Shaevitz, Nucl. Phys. B (Proc. Suppl.)  {\bf{19}} (1991) 270;
S.A. Rabinowitz et al., Phys. Rev. Lett. {\bf{70}} (1993) 134;
A.O. Bazarko et al., Z. Phys. {\bf{C 65}} (1995) 189.

\bibitem{Stron91}
B. Strongin et al., Phys. Rev. {\bf{D 43}} (1991) 2778.

\bibitem{CTEQ}
H.L. Lai et al., Phys. Rev. {\bf{D 51}} (1995) 4763; 
Phys. Rev. {\bf D 55} (1997) 1280.


\nonfrenchspacing
\end{thebibliography}
\end{document}